\def\BEQ{\begin{eqnarray}}
\def\EEQ{\end{eqnarray}}
\def\BML{\begin{mathletters}}
\def\EML{\end{mathletters}}
\def\BE{\begin{equation}}
\def\EE{\end{equation}}
\def\NN{\nonumber}

\let\d=\partial
\let\t=\theta
\let\La=\Lambda

\documentstyle[preprint,prb,aps,amsfonts,amssymb]{revtex}
\input epsf.tex
\draft
\begin{document}
\title{The Fermi Liquid as a Renormalization Group Fixed Point: the Role of
Interference in the Landau Channel}
\author{Gennady Y. Chitov and David S\'en\'echal}
\address{ Centre de Recherche en Physique du Solide et D\'epartement de
Physique,}
\address{Universit\'e de Sherbrooke, Sherbrooke, Qu\'ebec, Canada J1K 2R1.}
\maketitle
\begin{abstract}
\baselineskip=12pt
We apply the finite-temperature renormalization-group (RG) to a model based
on an effective action with a short-range repulsive interaction and a
rotation invariant Fermi surface. The basic quantities of
Fermi liquid theory, the Landau function and the scattering vertex, are
calculated as fixed points of the RG flow in terms of the effective action's
interaction function. The classic derivations of Fermi liquid theory, which
apply the Bethe-Salpeter equation and amount to summing direct particle-hole
ladder diagrams, neglect the zero-angle singularity in the exchange
particle-hole loop. As a consequence, the antisymmetry of the forward
scattering vertex is not guaranteed and the amplitude sum rule must be
imposed by hand on the components of the Landau function. We show that the
strong interference of the direct and exchange processes of particle-hole
scattering near zero angle invalidates the ladder approximation in this
region, resulting in temperature-dependent narrow-angle anomalies in the
Landau function and scattering vertex. In this RG approach the Pauli
principle is automatically satisfied. The consequences of the RG corrections
on Fermi liquid theory are discussed. In particular, we show that the
amplitude sum rule is not valid.
\end{abstract}

\baselineskip=12pt
\section{Introduction}

In 1956-1957 L.D. Landau formulated his theory of Fermi 
liquids.\cite{Landau57} The original phenomenological formulation of this
theory is based on an expansion near the ground state of the energy
functional in terms of variations of the distribution function (bosonic
variables). Later, Pomeranchuk derived the thermodynamic stability
conditions for this functional.\cite{Pomeran} Much effort has been
dedicated, including by Landau himself,\cite{Landau59} to vindicate some
intuitive assumptions of Landau and elucidate the foundations of the
phenomenological Fermi Liquid Theory (FLT). The field-theoretic
interpretation of the Landau FLT has reformulated the key notions and basic
results of the phenomenological theory entirely in terms of the fermionic
Green functions technique.\cite{Landau59,Lutt60,AGD,Nozieres} The
demonstration of the equivalence of the field-theoretic results obtained 
from the solution of the Bethe-Salpeter equation with the results obtained 
from the functional expansion and from the Boltzmann transport equation 
describing the collective modes, has become a textbook 
topic.\cite{AGD,Nozieres,Lifshitz80,PinesNoz,Baym91} The field-theoretic 
approach provided not only a solid basis to phenomenology, but also a 
potentially efficient method to calculate the phenomenological parameters of 
FLT from first principles. 

Current interest in non-Fermi Liquids in $d>1$ inspired a new wave of efforts
aimed at clarifying the foundations of the Landau FLT and the mechanisms of
its breakdown. Let us mention only two approaches, which can be seen as
sophisticated modern counterparts of the two classic formulations of the
Landau FLT. A bosonized treatment of Fermi liquids has recently been
developed\cite{Boson} in the framework of Haldane's formulation of
higher-dimensional bosonization.\cite{Haldane} At about the same time, the
Renormalization Group (RG) technique has been applied to interacting fermions
in $d>1$ with models based on fermionic field effective actions (see Refs
\onlinecite
{Benfatto,Shankar91,Polchinski93,Shankar94,Weinberg,Nayak,Hewson,GD95,NG96} 
and references therein). It both approaches it has been established, for
models with reasonable fermion-fermion effective interactions, that the Fermi
liquid phase is stable, whereas adding gauge-field interactions may drive
the system towards a Non-Fermi-Liquid regime, or may result in a Marginal
Fermi Liquid phase, like for composite fermions at the half-filled Landau
level.

The RG analysis of FLT presented here and in our previous work,\cite{GD95}
like other such analyses already published, starts from a low-energy
effective action with a marginal (in the RG sense) short-range interaction.
However, contrary to other works on the subject, our finite-temperature
RG approach revealed that, in the Landau channel of nearly forward
scattering quasiparticles, the effective interaction flows with successive
mode eliminations towards the Fermi surface, even in the absence of singular
or gauge interactions. In other words, the action's interaction (coupling
function) does not stay as a purely marginal under the RG transformation,
since its $\beta$-function is not identically zero. From the RG flow equations
the standard FLT results have been recovered.\cite{GD95}

It was also pointed out, and elaborated later in more detail by one of us
together with N. Dupuis in Ref.\onlinecite{NG96}, that the bare interaction
function of the low-energy fermion effective action cannot be identified with
the Landau interaction function. The latter, along with other observable
parameters of a Fermi liquid, should be calculated as a fixed point of the RG
equations.\cite{NG96} Let us briefly give two arguments for this. First,
identifying the Landau function with the effective action's bare interaction
is inconsistent with other standard FLT results, due to the role of Fermi
statistics. Indeed, in a stable Fermi liquid, the well-known relationship
between components of the scattering amplitude ($\Gamma_l$) and of the Landau
interaction function ($F_l$), i.e., 
$\Gamma_l= F_l/(1+F_l)$, cannot satisfy the Pauli principle for the
amplitude (the amplitude sum rule) if
$F$ has the symmetry properties of the action's bare interaction. (For the
explanation of this point see Sect.~\ref{deficientS} below). Second,
identifying the Landau function with the bare interaction is inconsistent
with the low-energy effective action method itself, in the way it is applied
to condensed matter problems. Namely, at the starting point of the
analysis, the bare parameters of the effective action, including the
interaction, are regular functions of their
variables.\cite{Polchinski93,Shankar94} It is known, however, that this is not
the case even for parameters of a normal Fermi liquid. For instance, the
scattering amplitude and the Landau function are two distinct limits of the
four-point vertex in the Landau channel when energy-momentum transfer goes to
zero. The non-analyticity of the forward scattering vertex appears in its
dependence both on the small energy-momentum transfer and, due to the
antisymmetry (crossing symmetry), on the small angles between incoming
(outgoing) particles lying near the Fermi surface. This contradiction
becomes flagrant if one couples the fermionic action with gauge fields since,
as shown by other methods,\cite{Stern} the Landau function for the marginal
Fermi Liquid of composite fermions at the half-filled Landau level develops a
delta-function singularity in the forward direction ($\t=0$). Such behavior
of the Landau function is related to the divergence of the quasiparticle's
effective mass, according to the theory of Halperin, Lee and Read for the
half-filled Landau level\cite{HLR} (see also Ref\onlinecite{Kopietz}). So,
coming back to our arguments, the Landau function cannot be a regular
interaction in the effective action at the starting point of the RG analysis.

The aim of the present study is twofold. Once the classic FLT results have
been recovered by the RG approach,\cite{GD95} the latter would loose its
appeal if it did not provide a constructive method for calculating the Fermi
liquid's parameters. This is especially important goal in the long-term
prospective of applying this powerful method to more complex strongly
correlated fermion systems. In this work we explicitly derive the Landau
function and the forward scattering vertex from the short-range effective
bare interaction. We do it in the one-loop RG approximation which takes into
account contributions of the direct ($ZS$) and exchange $(ZS')$ graphs. This
enables us to reveal singular features of the Landau function and scattering
vertex in the forward direction ($\theta =0$).

An equally important goal of this work is to resolve the old problem of FLT
with the Pauli principle. In its treatment of FLT, the field-theoretic
approach encountered a very subtle problem caused by Fermi statistics of
one-particle excitations and by the necessity to provide both stability for
the Fermi liquid and a solution for the two-particle vertex that meets the
Pauli principle.\cite{Mermin67,Lifshitz80} The problem was ``settled'' by
imposing the amplitude sum rule on the components of the Landau
quasiparticle's interaction function. The phenomenological FLT is spared from 
this problem partially by the way it is formulated, partially because it says 
nothing about the quasiparticle scattering amplitudes. (A detailed discussion
of this problem, which lies at the heart of the present study, is postponed 
until Sec.\ref{deficientS}, where it will be put in contact with the present 
RG approach.) The same problem arose in our previous work\cite{GD95} in the
form of a ``naturalness problem''\cite{Polchinski93} of the effective action:
the effective action had to be ``fine tuned'' in order for the scattering
amplitude to meet the Pauli principle. We will show that if quantum
interference of the direct and exchange processes is taken into account, this
problem is eliminated in a natural manner.

The paper is organized as follows. Sections \ref{modelS} and \ref{couplingS}
are introductory: we define the effective action of the model and the
coupling functions (the bare interaction) and vertices to be calculated in
the Landau interaction channel. In Section \ref{RGeqS}, which is rather
technical, the one-loop RG equations for the two-dimensional case are
derived. Section \ref{deficientS} explains some of the weak points of the
standard FLT results and argues for their partial revision. In Section
\ref{solutionS} we give a numerical and approximate analytical solution of
the coupled RG equations for spinless fermions. In Section
\ref{discussionS} we present and discuss our results for the Landau function
and the scattering vertex calculated at different temperatures. In
Section~\ref{contactS} we relate this study to the standard treatment of
Fermi liquid Theory. The consequences of the RG corrections on FLT results
are discussed.

\section{The Model}\label{modelS}

We apply the Wilson-Kadanoff renormalization scheme in the framework
developed earlier for a model with $SU(N)$-invariant short-range effective
interaction and rotation invariant Fermi surface in spatial dimensions
$d\leq3$ at finite temperature.\cite{GD95} In order to make the discussion as
clear as possible, we concentrate in this work on $2D$
spinless ($N=1$) fermions. This simple model has nevertheless all the
necessary qualities to illustrate our key points and to demonstrate the new
features brought by the RG analysis of a Fermi liquid. In this case the RG
equations take their simplest form, since only the antisymmetric
momentum-frequency dependent parts of the interaction and vertices are present
(they were labeled by A in Ref.\onlinecite{GD95}).

The partition function in terms of Grassmann variables
is given by the path integral
\BE\label{Z} 
Z = \int {\cal D}\bar\psi{\cal D}\psi~e^{S_0 + S_{\rm int}}\EE
wherein the free part of the
effective action\cite{Polchinski93,Shankar94,Weinberg} is
\BE\label{S0}
S_0 =\int_{({\bf 1})}\bar\psi({\bf 1})\left[
i\omega_1+\mu-\epsilon({\bf K}_1)\right]\psi({\bf 1}) ~.\EE
We introduced the following notation:
\BML
\label{notation}
\BE \int_{({\bf i})}~~\equiv~~{1\over\beta}\int{d{\bf
K}_i\over(2\pi)^2}\sum_{\omega_i}\EE
\BE
({\bf i})~~\equiv~~({\bf K}_i,\omega_i),\EE 
\EML
where $\beta$ is the inverse temperature, $\mu$ the chemical
potential,
$\omega_i$ the fermion Matsubara frequencies. We set $k_B=\hbar=1$. 
The interacting part of the action is
\BE
\label{Sint}
S_{\rm int} = -\frac14\int_{({\bf 1},{\bf 2},{\bf 3},{\bf 4})}
\bar\psi({\bf 1})\bar\psi({\bf 2})\psi({\bf 3})\psi({\bf 4}) 
\Gamma^{\La_0}({\bf 1},{\bf 2};{\bf 3},{\bf 4}) 
\beta(2\pi)^2 \delta({\bf 1}+{\bf 2}-{\bf 3}-{\bf 4})~,\EE
where $\delta(\cdots)$ stands for a Dirac delta function for the momenta
and a Kronecker delta for the Matsubara frequencies.
The function $\Gamma^{\La_0}$ is antisymmetric under the exchange $({\bf
1} \leftrightarrow {\bf 2})$ and $({\bf 3} \leftrightarrow {\bf 4})$. The bare
cutoff $\La_0$ of the action is introduced such that each vector
${\bf K}_i$ in the effective action lies in a shell of thickness $2\La_0$
around the Fermi surface. We denote this shell, i.e., the support of the
effective action in the momentum space, as $C^2_{\La_0}$. The Matsubara
frequencies are allowed to run over all available values. We presume that the
density of particles in the system is kept fixed. 

The one-particle excitations are linearized near the Fermi surface, and
therefore the bare one-particle Green's function for the free part of action
$S_0$ is:
\BE\label{G0}
G_0^{-1}({\bf K}_1,\omega_1) = i\omega_1+\mu-\epsilon({\bf K}_1)
\approx i\omega_1-v_F(K_1-K_F) \equiv i\omega_1-v_Fk_1 ~.\EE
In the integration measure only the relevant part is kept:
\BE\label{measure}
\int d{\bf K}~=~\int_{-\La_0}^{\La_0}
\int_0^{2\pi}{(K_F+k)}dkd\theta ~\approx~
K_F\int_{-\La_0}^{\La_0}\int_0^{2\pi} dkd\theta \EE
The temperature $T$ is restricted by the condition 
\BE\label{TCond} T\ll ~v_F\La_0 ~.\EE 

The relevant physical information can be obtained by studying interactions
of particles scattering with small momentum and energy transfer (we call it
the {\it Landau channel}), and those with nearly opposite incoming (outgoing)
momenta (the BCS channel). Since we are interested in the repulsive case,
we presume that stability conditions against Cooper pairing are fulfilled,
and we concentrate on the Landau channel.

\section{Coupling functions and vertices in the Landau channel}
\label{couplingS}

Let us clarify the meaning of the quantities entering the effective action.
Consider the vertex function 
$\Gamma({\bf 1},{\bf 2};{\bf 3},{\bf 4})$, constructed from the connected 
two-particle Green's function
${ G}_2^c({\bf 1},{\bf 2};{\bf 3},{\bf 4}) = 
-\langle\psi({\bf 1})\psi({\bf 2})
\bar\psi({\bf 3})\bar\psi({\bf 4})\rangle_c$
by amputation of the external legs. Here $\langle ...\rangle$ means an average
with the effective action (\ref{S0},\ref{Sint}) which contains only ``slow''
modes, lying in the support $C^2_{\La_0}$. Once auxiliary source fields
(with momenta inside the shell $C^2_{\La_0}$) coupled to the action's
Grassmann fields $\{\psi$, $\bar\psi \}$ have been introduced, such connected
$n$-particle Green's functions can be defined as functional derivatives of the
source-dependent generating functional.\cite{Metzner}
At tree-level, $\Gamma({\bf 1},{\bf 2};{\bf 3},{\bf 4})|_{\rm tree} = 
\Gamma^{\La_0}({\bf 1},{\bf 2};{\bf 3},{\bf 4})$.
The bare vertex $\Gamma^{\La_0}$ (in the sense of the effective action
(\ref{S0},\ref{Sint})) can be defined in the same fashion as $\Gamma$, with 
the difference that $\Gamma^{\La_0}$ is the result of averaging over 
the ``fast'' modes (those {\it outside} $C^2_{\La_0}$) with the
{\it microscopic} action. Contrary to
$\Gamma$, the vertex $\Gamma^{\La_0}$ is not a physical 
observable, since it is not the result of an integration over all degrees of
freedom.

Taking into account momentum and frequency conservation, we use the
following notation for the nearly forward scattering vertex:
\BE\label{GammaTr}
\Gamma({\bf 1},{\bf 2};{\bf 1}+{\cal Q},{\bf 2}-{\cal Q})
\equiv \Gamma({\bf 1},{\bf 2};{\cal Q}) ~,\EE
with the transfer vector 
\BE\label{Transfer}
{\cal Q} = {\bf 3}-{\bf 1} \equiv ({\bf Q},\Omega ) \EE
such that $Q \ll K_F$ ($\Omega $ is a bosonic Matsubara frequency).
We write the momentum ${\bf K}_i$ as ${\bf K}_i={\bf K}_F^i+{\bf k}_i$
where ${\bf K}_F^i$ lies on the Fermi surface and ${\bf k}_i$
($|{\bf k}_i| \leq \La_0$) is normal to the Fermi surface at the point 
${\bf K}_F^i$.

In order to calculate physical quantities, we must perform an average 
with the effective action (\ref{S0},\ref{Sint}), i.e., we must integrate out 
the ``slow'' modes, which lie inside $C^2_{\Lambda_0}$, in the corresponding 
path integrals. This is done in Wilson's RG approach
by successively integrating the high-energy modes in
$C^2_{\La_0}$, i.e., by progressively reducing the momentum cutoff from
$\La_0$ to zero. We define a RG flow parameter $t$ such that the cutoff
at an intermediate step is $\La(t)=\La_0 e^{-t}$. Integrating over
the modes located between the cutoffs $\La(t)$ and $\La(t+dt)$, a
recursion relation (in the form of a differential equation) can be found for
the various parameters of the action.
This equation (or set of equations) is then solved from $t=0$ to $t\to\infty$
and this yields the fixed-point value of the parameters of the
action. The physical quantities are then obtained from these parameters, e.g.
by functional differentiation if they are source fields. 

A considerable simplification of this problem comes 
from the scaling analysis of the low-energy effective action
using the smallness of the scale $\La/K_F$.\cite{Shankar94} 
A tree-level analysis shows that the only part of the coupling function
$\Gamma^\La$ which is not irrelevant couples two incoming and two outgoing
particles with the same pairs of momenta $({\bf K}_F^1,{\bf K}_F^2)$ lying on
the Fermi surface. The dependence of the coupling function on ${\bf k}_i$ and
on the frequencies $\omega_i$ is irrelevant and can be omitted.
When the initial cutoff $\La_0$ satisfies
condition (\ref{TCond}), we can unambiguously define a {\it bare} 
coupling function which depends {\it only} on the angle between the
incoming (or outgoing) momenta. This bare coupling function 
is given by the vertex $\Gamma^{\La_0}({\bf 1},{\bf 2};{\cal Q})$
in the zero transfer limit (${\cal Q}=0$) where the two external momenta are 
put on the Fermi surface and the external frequencies are
$\omega_{\rm min}\equiv\pi T$ (the latter will be dropped from now on). 
\BE\label{Fdef}
U({\bf K}_F^1,{\bf K}_F^2) \equiv \frac12\nu_F
\Gamma^{\La_0}({\bf K}_F^1,{\bf K}_F^2; 0) ~,\EE
where $\nu_F = K_F/ \pi v_F$ is the free density of states
at the Fermi level. Each vector 
${\bf K}_F^i$ may be specified by a plane polar angle $\t_i$.
The function $U$ is an even function of the relative angle $\t_{12}$
between ${\bf K}_F^1$ and ${\bf K}_F^2$. The only remnant of the 
antisymmetry of $\Gamma^{\La_0}$ (the Pauli principle) is the 
condition:\cite{GD95}
\BE\label{Pauli} U(0)=0 ~.\EE

As shown earlier,\cite{GD95} the tree-level picture
becomes more complicated when we carry out the mode elimination inside
$C_\La^2$. It turns
out that simply discarding the frequency dependence of $\Gamma^{\La}$
and identifying the momenta ${\bf K}_F^1 \rightleftharpoons {\bf K}_F^3$, 
${\bf K}_F^2 \rightleftharpoons {\bf K}_F^4$ is an ill-defined procedure
when the running cutoff $\La$ becomes of the order of the temperature
($v_F \La \sim T$). The ambiguity arises when calculating the one
loop-contribution from, say, the $ZS$ graph, since this contribution is not an
analytic function of the transfer $\cal Q$ at
${\cal Q}=0$.\cite{AGD,Lifshitz80,Mermin67}
To describe correctly the parameters of the Fermi liquid, one should retain 
the dependence of the coupling function 
$\Gamma^{\La}({\bf K}_F^1,{\bf K}_F^2;{\cal Q})$
on the energy-momentum transfer ${\cal Q}$. Retaining this
${\cal Q}$-dependence allows the calculation of response functions or
collective modes of the Fermi liquid.\cite{Dupuis96} For the purpose of the
present study we define two coupling functions ($\Gamma^Q$ and $\Gamma^\Omega
$), depending on the order in which the limits of zero momentum- ($\bf Q$) 
and energy-transfer ($\Omega$) are taken:
\BML
\label{two-limits}
\BE
\Gamma^Q(\t_{12}) = \lim_{Q\to0}\Bigl 
\lbrack \Gamma(\t_{12},{\cal Q}) 
\Bigl \vert_{\Omega =0} \Bigr \rbrack \,, \EE
\BE
\Gamma^\Omega(\t_{12}) = \lim_{\Omega\to0}
\Bigl\lbrack \Gamma(\t_{12},{\cal Q}) \Bigl
\vert_{Q=0} \Bigr \rbrack \EE
\EML
We use dimensionless vertices, by including in their definition the 
factor $\frac12 \nu_F$, like in Eq.~(\ref{Fdef}).
The functions $\Gamma^{Q, \Omega}(\t)$ are even functions of the angle $\t$.
We will not explicitly indicate their dependence on the cutoff
$\La$, unless necessary. We will indiscriminately call these functions
(running) vertices.

Let us summarize: The effective action is defined on the support
$C^2_{\La_0}$ with the bare coupling function $\Gamma^{\La_0}$, which is
presumably an analytic function of its variables and is marginal at tree
level. While performing the mode elimination within $C^2_{\La_0}$, we
need to calculate the flow of the two vertices
$\Gamma^Q$ and $\Gamma^\Omega$. The bare coupling $\Gamma^{\La_0}$
has an unambiguous meaning only as the common initial point of the RG flow 
trajectories of $\Gamma^Q$ and $\Gamma^\Omega$. The fixed point values
$\Gamma^{Q*} \equiv \Gamma^Q(t=\infty)$ and 
$\Gamma^{\Omega *} \equiv \Gamma^\Omega(t=\infty)$ are physical observables:
the first one is the $Q$-limit of the vertex $\Gamma$ (as defined at the
beginning of this section) and is the scattering amplitude of
quasiparticles with all four external momenta lying on the Fermi surface. The
second one is the unphysical limit ($\Omega$-limit) of the vertex $\Gamma$ and
is identified with the Landau function.\cite{Landau59}

\section{The RG equations in the Landau channel}
\label{RGeqS}

There are three
Feynman diagrams contributing to the RG flow at the one-loop level (see
Fig.~1), denoted $ZS$ (zero sound), $ZS'$ (Peierls), and $BCS$. 
The $BCS$ graph
contribution preserves the antisymmetry of the vertex,
while those of the $ZS$ and $ZS'$ graphs separately do not: only their
combined contribution ($ZS+ZS'$) is antisymmetric under exchange of
incoming (or outgoing) particles. To respect the Pauli principle, it is
therefore necessary to take into account both the $ZS$ and $ZS'$
contributions to the RG flow. In this work we discard the symmetry-preserving
contribution of the $BCS$ graph to the RG flow of the vertices in the Landau
channel. Thus, we leave out the interference near $\t=\pi$ of the Landau
channel with the BCS channel, which leads to the Kohn-Luttinger
effect.\cite{Shankar94}

The formal analytic expression of the $ZS$ graph is
\BE\label{ZS}
ZS~=~-\int_{({\bf 5})}\Gamma({\bf 1}, {\bf 5};{\bf
1}+{\cal Q},{\bf 5}-{\cal Q}) 
\Gamma({\bf 5}-{\cal Q},{\bf 2};{\bf 5},{\bf 2}-{\cal Q}) G({\bf
5})G({\bf 5}-{\cal Q}) ~,\EE
wherein the transfer vector ${\cal Q}$ is given by (\ref{Transfer}).
To calculate the contribution of this graph to the RG flow of
$\Gamma^Q$ and $\Gamma^\Omega$, we only need to keep 
the dependence on the momenta ${\bf K}_F^i$ and on the transfer 
${\cal Q}$ in the vertices on the r.h.s. of (\ref{ZS}).
Momentum and energy conservation is already taken
into account in (\ref{ZS}). The phase space restrictions are satisfied
automatically for any ${\bf K}_5 \in C^2_{\La}$ in the limit ${\cal Q}
\to 0$. When
${\bf K}_1$ and ${\bf K}_2$ lie on the Fermi surface and ${\cal Q} \to 0$,
the r.h.s. of (\ref{ZS}) contains both vertices of type (\ref{two-limits})
with ${\bf K}_F^5$ running freely around the Fermi surface during the angular
integration. Thus, for this graph, all the phase space is available for
integration. The summation over $\omega_5$ of the Green's functions product
on the r.h.s. of (\ref{ZS}) when ${\cal Q} \to 0$ gives zero in the
$\Omega$-limit, and thus
\BE\label{GammaOmZS}
{{\d \Gamma^\Omega} (\t_1 -\t_2) 
\over {\d t}}
 \Biggl \vert_{\rm ZS}=0 ~.\EE
The ${\bf Q}$-limit of the same product gives a factor 
$\frac14 \beta \cosh^{-2}(\beta v_Fk_5/2)$, and accordingly\cite{GD95}
\BE\label{GammaQZS}
{\d\Gamma^Q(\t_1 -\t_2)\over\d t}
\Biggl \vert_{\rm ZS}= {\beta_R \over \cosh^2(\beta_R)}
\int_{- \pi}^\pi {{d\t } \over {2\pi }} 
\Gamma^Q (\t_1-\t )
\Gamma^Q(\t -\t_2) ~,\EE
where we introduced a dimensionless temperature flow parameter:
\BE\label{Br}
\beta_R(t) \equiv \frac12 v_F \beta \La (t) ~.\EE

We now turn our attention to the $ZS'$ graph. Its analytic form is
\BE\label{ZS'}
ZS'~=~~\int_{({\bf 5})}
\Gamma({\bf 1}, {\bf 5};{\bf 1}+{\cal Q}',{\bf 5}-{\cal
Q}') 
\Gamma({\bf 5}-{\cal Q}',{\bf 2};{\bf 5},{\bf 2}-{\cal
Q}') G({\bf 5})G({\bf 5}-{\cal Q}') ~,\EE 
wherein ${\cal Q}' \equiv {\bf 2}-{\bf 1}-{\cal Q}$
can be thought of as an ``effective'' transfer vector for this graph. 
For $|{\bf K}_2-{\bf K}_1| \not=0$ the limit ${\cal Q} \to 0$ of the
r.h.s. of (\ref{ZS'}) is single-valued and equivalent to the
$Q$-limit.\cite{Mermin67} The Green's functions contribution to this graph is
\BE\label{GProduct}
{1\over\beta}\sum_{\omega_5} G({\bf 5})G({\bf 5}-{\cal Q}') 
\Biggl\vert_{{\cal Q} = 0}= -{1\over 2} {{\tanh \Bigl \lbrack {\beta
\over 2}(\epsilon ({\bf K}_5)-\mu) \Bigr \rbrack - \tanh \Bigl \lbrack
{\beta\over2} (\epsilon ({\bf K}_5-{\bf K}_2-{\bf K}_1)-\mu) \Bigr \rbrack} 
\over {\epsilon ({\bf K}_5) -\epsilon ({\bf K}_5-{\bf K}_2-{\bf K}_1)} }~.\EE
If $|\t_1-\t_2| \ll {T /v_F K_F}$ the
r.h.s. of Eq.(\ref{GProduct}) becomes $-\frac14 \beta \cosh^{-2}(\beta
v_Fk_5/2)$. The calculation of the $ZS'$ contribution is more subtle, since
even in the zero-transfer limit ${\cal Q} \to 0$ (in any order), the
vector ${\bf Q}' \vert_{{\cal Q} \to 0} = {\bf K}_2-{\bf
K}_1$ is free to take any modulus in the interval 
$\lbrack 0, 2K_F \rbrack$ as the angle $\t_1-\t_2$ varies. A large 
${\bf Q}'$ kicks the vertex momenta on the r.h.s. of (\ref{ZS'})
outside of $C^2_{\La}$, even if ${\bf K}_5 \in C^2_{\La}$. In such
cases the contribution of the $ZS'$ graph is cut off, except for special
positions of the vector ${\bf K}_5$ running over the Fermi surface. Thus,
for an arbitrary angle $\t_1-\t_2$, not all the phase space is available
for integration.

To understand where this elimination of the $ZS'$ contribution comes from, we
must keep in mind that our effective action has support
$C^2_{\La}$ in momentum space. Let us consider the $ZS'$ graph (see
Fig.~1) when all external momenta satisfy momentum
conservation and lie in $C^2_{\La}$. It suffices then to check whether
the internal momenta (${\bf K}_5$ and ${\bf K}_5-{\bf Q}'$) lie in
$C^2_{\La}$ when ${\bf K}_F^5$ runs around the Fermi surface during the
integration. From Fig.~2A we see that if $|{\bf Q}'| >2 \La$, the
loop momenta lie both in $C^2_{\La}$ only at special values
of ${\bf K}_F^5$ (the shaded regions), i.e., only small fragments of
phase space are available for integration. At smaller 
${\bf Q}'$ (cf. Fig.~2B) these intersections form a connected region
and ${\bf K}_F^5$ is free to run around the Fermi surface.
If we completely neglect the
$ZS'$ graph when the intersection is disconnected (in Fig.~2A), the
contribution of this graph to the RG flow at $|{\bf Q}'| < 2 \La$ is
calculated in the same way as that of the $ZS$ graph. Since $|{\bf
K}_1|=|{\bf K}_2|=K_F$ and 
${\bf Q}' \vert_{{\cal Q} \to 0} = {\bf K}_2-{\bf K}_1$, the
condition $|{\bf Q}'| < 2 \La$ is equivalent to the condition
$|\sin((\t_1-\t_2)/2)| < \La / K_F$ for the angle between 
${\bf K}_1$ and ${\bf K}_2$.

Taking into account both the contributions of the ZS and ZS$'$ graphs, the RG
equations for $\Gamma^{Q,\Omega }$ can be written in 
{\it implicit form}:\cite{NG96}
\BE
{\d\Gamma^Q \over\d t} = 
{\d\Gamma^Q \over\d t}\Biggl \vert_{\rm ZS} +
{\d\Gamma^Q \over\d t}\Biggl \vert_{\rm ZS'}\EE
\BE
{\d\Gamma^\Omega \over \d t} = 
{\d\Gamma^\Omega \over \d t}\Biggl\vert_{\rm ZS'} = 
{\d\Gamma^Q \over\d t}\Biggl \vert_{\rm ZS'} ~.\EE
Summing up all formulas, we obtain the following system of RG equations:
\BML
\label{RGangle}
\BE
{\d\Gamma^Q(2 \phi)\over\d t} = {\beta_R \over\cosh^2\beta_R}
\int_{-\pi}^\pi {d\t\over 2\pi} \Gamma^Q(\phi-\t)
\Gamma^Q(\t + \phi)
+{\d\Gamma^\Omega(2\phi)\over\d t}\EE
\BE {\d\Gamma^\Omega(2\phi)\over\d t} = 
-\beta_R \Theta(\t_c- |\phi|) 
\int_{-\pi}^\pi {d\t\over2\pi} 
\Gamma^Q(\phi-\t)
\Gamma^Q(\t +\phi) Y(\phi,\t;\beta_R)~.\EE
\EML
To simplify those formulas we parametrized the angular dependence of the
vertices in Eqs.~(\ref{RGangle}) by the angle $\phi$ between ${\bf
K}_F^1$ and $({\bf K}_F^1+{\bf K}_F^2) $, $| \phi| \in [0, \pi/2]$. 
The small $ZS'$ contribution coming from $|\sin\phi|>\La(t)/K_F$ (Fig.
2A) was neglected, which is accounted for by the Heaviside step function
$\Theta$, wherein $\t_c \equiv \arcsin( \La(t) /K_F)$.
We also defined the function
\BE
\label{Ybetax}
Y(\phi, \t; \beta_R) \equiv {1 \over \beta_{Q^\prime}}
{\sinh(2\beta_{Q^\prime})\over{\cosh(2\beta_R) + 
\cosh(2\beta_{Q^\prime})}}\EE
\BE
\beta_{Q^\prime}\equiv\beta_F\sin\t\sin\phi,~~
\beta_F\equiv\beta v_F K_F ~.\EE
which arises in the calculation of the $ZS'$ contribution (\ref{GProduct}).
Notice that
\BE\label{limY}
\lim_{\beta_{Q^\prime}\to 0} Y(\phi,\t;\beta_R) = {1\over\cosh^2\beta_R}~.\EE 
{}From Eqs.~(\ref{RGangle}a,\ref{Ybetax},\ref{limY}) we see that 
at small angles $(|\phi|\lesssim {T /v_F K_F})$ there is a strong
interference between the $ZS$ and $ZS'$ contributions. This
interference depletes the RG flow of
$\Gamma^Q (\phi)$ at small angles. Moreover, at
$\phi=0$ the flow is exactly zero, for the two contributions have the same
thermal factor $\beta_R\cosh^{-2}(\beta_R)$:\cite{note0}
\BE\label{Gamma0}
{\d\Gamma^Q (\phi=0, t) \over\d t} =
0~,~\forall~t~.\EE

The initial conditions for the flow equations (\ref{RGangle}) are:
\BE\label{Init}
\Gamma^Q (\phi, t=0)= \Gamma^\Omega
(\phi, t=0)=U(\phi)~.\EE
Recall that the fixed points $\Gamma^{Q*}$ and $\Gamma^{\Omega *}$ of the
vertices $\Gamma^Q$ and $\Gamma^\Omega$ are the forward scattering vertex and
the Landau interaction function, respectively.
From Eqs.~(\ref{Gamma0},\ref{Init},\ref{Pauli}) we conclude that the RG
equations for the forward scattering vertex preserve the Pauli principle at
any point of the RG flow trajectory
\BE\label{GPauli}
\Gamma^Q (\phi=0 , t)=0~,~~ \forall ~ t~,\EE
while the ``uncompensated'' RG flow generated by the $ZS'$ graph drives the
vertex $\Gamma^\Omega$ to a fixed point value (the Landau function), which 
does not satisfy the Pauli principle, i.e., $\Gamma^{\Omega*}(\phi=0) \not=0$. 

\section{Deficiencies of the decoupled approximations in the Landau channel}
\label{deficientS}

Before finding a solution (exact or approximate) to the flow equations
(\ref{RGangle}) which fully takes into account the coupling of $\Gamma^Q$
and $\Gamma^\Omega$, we will comment on approximate solutions in which this
coupling is neglected. The Landau channel, as defined in this paper, includes,
at one-loop RG, both the direct ($ZS$) and exchange ($ZS'$)
quasiparticle-quasihole loops with a small transfer $\cal Q$. We will call
{\it decoupled} any treatment of the Landau channel which does not
explicitly take into account both the direct and exchange contributions. It
is shown below that solutions for the forward scattering vertex provided by
decoupled methods fail to meet the requirements of the Fermi statistics.
Tackling the Pauli principle by imposing additional constraints on the
solutions (sum rules) leads to conceptual difficulties discussed below. 

To shorten notation we drop 
upper labels ($ Q, \Omega$), and define $\Gamma$ ($F$) as the running
vertex whose fixed point is the forward scattering vertex (resp. the Landau
function).

Let us first solve the RG equations in the decoupled approximation.
If we neglect completely the $ZS'$ contribution in Eqs.~(\ref{RGangle}) and
perform a Fourier transformation, we recover a familiar system of
equations,\cite{GD95} with its RPA-like solution in which all harmonics are
decoupled:
\BML
\label{RGRPA}
\BE
{\d\Gamma_l\over\d\tau} = \Gamma_l^2 ~~\Longrightarrow~~ 
\Gamma_l^{\rm RPA} (\tau)= {\Gamma_l(\tau_0)\over 1+(\tau_0-\tau)
\Gamma_l(\tau_0)} \EE
\BE
{\d F_l\over\d\tau} =0~~~\Longrightarrow~~F_l^{\rm RPA} (\tau)= {\rm cst} ~.\EE
\EML 
We introduced the auxiliary parameter
$\tau \equiv \tanh \beta_R$ ($\tau \in [0, \tau_0]$), with 
$\tau_0 \equiv\tanh \beta_0$ and 
$\beta_0 \equiv \frac12 v_F \beta \La_0$. Since the temperature in the
effective action is restricted by the condition (\ref{TCond}), we can set
$\tau_0=1$ for all practical purposes. 

With the initial conditions $\Gamma_l(\tau_0)=F_l(\tau_0)=U_l$
(cf Eq.~(\ref{Init})), the fixed points of Eqs.(\ref{RGRPA}) are
\BE
\label{RPAFP}
{\rm (a)}~~~\Gamma_l^* = {U_l\over 1+U_l} \qquad\qquad
{\rm (b)}~~~F_l^* = U_l~,\EE
with the following stability conditions for the fixed point:
\BE\label{Stab1}
U_l > -1~, ~~~\forall~ l~,\EE
which are the Stoner criteria well-known from the RPA approach.
The bare interaction satisfies the Pauli principle (cf. Eq.(\ref{Pauli}))
\BE\label{Pauli2}
\sum_{l=-\infty}^\infty U_l = 0 ~.\EE
If the vertex $\Gamma$ is to satisfy the Pauli principle, the
condition
\BE\label{SR}
\sum_{l=-\infty}^\infty {U_l\over 1+U_l} =0 \EE
must be imposed on the r.h.s. of (\ref{RPAFP}a). However, it has been
known for a long time that conditions (\ref{SR}) and (\ref{Pauli2}) are 
incompatible, unless the stability conditions (\ref{Stab1}) are
broken.\cite{Mermin67} Indeed, subtracting (\ref{Pauli2}) from (\ref{SR}), we
find 
\BE\label{BrSt}
\sum_{l=-\infty}^\infty {U_l^2 \over 1+U_l} =0 ~,\EE
which cannot be satisfied without violation of (\ref{Stab1}).

This proves that the antisymmetric bare interaction $U$ cannot be at the
same time a fixed point of the RG flow and the Landau function, unless the
classic FLT formulas are unapplicable. The accepted cure to this paradox
is to give up the Pauli principle on the Landau function,
because of the neglected $ZS'$ contribution.\cite{Mermin67} 
In the RG approach, this may be accomplished (in the decoupled
approximation) by letting the $ZS'$ contribution drive the bare interaction
$U$ towards the Landau function $F^*$ during an earlier stage of mode
elimination, and then by solving the RG equations (\ref{RGRPA}) with
$F^*$ as a new renormalized ``bare'' interaction.\cite{GD95,NG96} This leads
to the well-known relationship between the scattering vertex and the Landau
function 
\BE\label{RPALandau}
\Gamma_l^* = {F_l^* \over 1+F_l^*} \EE
Because of the $ZS'$ contribution, the Pauli principle does not
apply to $F^*$ (a), while it is enforced on the vertex $\Gamma^*$ through a
sum rule (b):
\BE\label{SRLandau}
{\rm (a)}~~\sum_{l=-\infty}^\infty F_l^*\neq 0 \qquad\qquad
{\rm (b)}~~\sum_{l=-\infty}^\infty {F_l^*\over 1+F_l^*} = 0~.\EE
In doing so, the stability conditions (\ref{Stab1}) are
modified as follows
\BE\label{StabPom} F_l^* >-1~, ~~~\forall~ l~,\EE
i.e., they become Pomeranchuk's stability conditions for the
Fermi liquid, originally obtained on thermodynamic grounds.\cite{Pomeran} 
Such a decoupled RG treatment of the direct and exchange loops makes
Eqs.~(\ref{SRLandau}) compatible with the conditions (\ref{StabPom}). 

However, the sum rule (\ref{SRLandau}b) is ``unnatural'', in the following
sense. The bare interaction can in principle be traced from a microscopic
Hamiltonian. For instance, let us consider the spinless extended Hubbard 
Hamiltonian on a square lattice (with lattice spacing $a$) at low filling, 
with nearest-neighbor repulsive interaction ($U^{\rm nn}$).
Fourier-transforming and antisymmetrizing the interaction, we end up with the
following coupling function of the microscopic Hamiltonian:
$U^A_{\rm mic} ({\bf K}_1,{\bf K}_2;{\bf K}_3,{\bf K}_4) ~\simeq~ -\frac14 a^2
U^{\rm nn}\cdot ({\bf K}_1-{\bf K}_2)\cdot 
({\bf K}_3-{\bf K}_4)$.\cite{Shankar94} Let us 
choose this interaction as a trial bare dimensionless coupling function:
\BE\label{Ubare}
U(\t_1-\t_2)={\cal U} \sin^2\bigg({\t_1-\t_2
\over 2}\bigg)~,\EE
wherein all parameters are hidden within a single
coefficient ${\cal U}$. The only nonzero Fourier components $U_l$ of the
interaction are:
\BE\label{UbareF}
U_0= \frac12 {\cal U}~~,~~U_{\pm1}=- \frac14 {\cal U}~.\EE
The interaction (\ref{Ubare}) satisfies the Pauli principle
(\ref{Pauli},\ref{Pauli2}). The RPA sum rule (\ref{SR}) imposes an additional
constraint, which the interaction (\ref{Ubare}) does not satisfy. 
If we suppose that the ``improved'' results (\ref{RPALandau},\ref{SRLandau}b)
are always true, then, starting from any kind of microscopic interaction
(e.g., the bare interaction (\ref{Ubare})) and integrating ``fast modes''
outside the immediate vicinity of the Fermi surface, we have to end up with a
``fine tuned'' interaction, for any interaction has to be ``fine tuned'' in
order to satisfy (\ref{SRLandau}b). The integral of the flow (\ref{GPauli})
(or, equivalently the sum rule (\ref{Gsumrule}) below) is not a fine tuning, 
since firstly, the bare interaction at the initial point can be always 
antisymmetrized, and, secondly, we have an exact cancellation of the RG 
flow for the vertex $\Gamma$ at zero angle due to direct and exchange 
contributions, thus preserving (\ref{GPauli}). On the contrary, there is no 
reason for 
any bare interaction to satisfy (\ref{SR}) at the beginning, nor is there
a mechanism to provide the fine tuning (\ref{SRLandau}b) on other parts of 
the RG trajectory.

These difficulties are not specific to the decoupled RG approximation, since
the latter is strictly equivalent to the diagrammatic microscopic derivation
of FLT\cite{Landau59,AGD,Lifshitz80} leading to the same results
(\ref{RPALandau},\ref{SRLandau},\ref{StabPom}). The decoupled RG treatment is
equivalent to applying the Bethe-Salpeter equation with the particle-hole $ZS$
loop singled out, $F$ being the vertex irreducible in this loop. There are no
{\it a priori} reasons in that approach to demand this vertex to satisfy the
Pauli principle. The rearrangment of diagram summations in the Bethe-Salpeter
equation leading to (\ref{RPALandau}) is based on the assumption that the
vertex irreducible in the direct particle-hole loop ($ZS$) is a regular
function of its variables, neglecting the zero-angle singularity (at
$T=0$\cite{note0}) in the $ZS'$ loop. As a consequence, the Pauli principle
for the scattering vertex ${ \Gamma}^*$ is not guaranteed in the final result
and ``the amplitude sum rule'' (\ref{SRLandau}b) must be imposed by hand. 
The solution (\ref{RPALandau}) of the Bethe-Salpeter equation is tantamount to
the summation of the ladder diagrams built up from the $ZS$ loops, wherein the
Landau function stands as the bare interaction. For this reason, the solution
(\ref{RPALandau}) we will call the ``{\it the
$ZS$-ladder approximation}'' in the following. We refer the reader to a paper
of A. Hewson\cite{Hewson} wherein a ``generalized'' Bethe-Salpeter equation
for Fermi liquids, which explicitly takes into account both the $ZS$ and $ZS'$
loops, is derived. For further discussion on this issue, see also
Ref.\onlinecite{Mermin67}.

\section{Solution of the coupled RG equations}
\label{solutionS}

\subsection{Exact Numerical Solution}

The coupled integro-differential flow equations (\ref{RGangle}) may be solved
numerically. The functions $\Gamma(\t)$ and $F(\t)$ are then defined on a
discrete grid of angles, and simple linear interpolation is used to represent
them between the grid points. The grid spacing is not uniform: it has to be
very small near $\t=0$, where the flow is singular, but may be larger
elsewhere. The RG equations then reduce to a large number of coupled
nonlinear differential equations, which are solved by a fourth-order
Runge-Kutta method with adaptive step-size. Typically, a grid of a few
hundred points is sufficient (we take advantage of the symmetry of the
functions). Of course, the numerical solution was checked to be
indistinguishable from the (exact) RPA solution when the $ZS'$ contribution
is discarded. 

An example of solution for the spinless case with the interaction function
(\ref{Ubare}) is shown on Fig. 3(A), at various temperatures. The interaction
function $U(\t)$ and the RPA solution $\Gamma^{\rm RPA}(\t)$ are also shown.
This solution will be discussed in Sect.~\ref{discussionS}.

\subsection{Approximate Analytical Solution}

The flow equations (\ref{RGangle}) may also be solved analytically, albeit
only approximately. In this section we give the approximate solution for 
the fixed points $\Gamma^*$ and $F^*$ both in terms of Fourier components 
and in terms of angular variables. (See Eqs.~(\ref{ApGam},\ref{Gamang})
below.)

The Fourier transform of Eqs.~(\ref{RGangle}) is
\BML\label{GFex}
\BE
{\d\Gamma_n\over\d\beta_R} = {1\over\cosh^2\beta_R}\Gamma_n^2 +
{\d F_n\over \d\beta_R} \EE
\BE {\d F_n \over \d \beta_R} 
= -\sum_{l,m=-\infty}^\infty {\cal Y}_{n-m,2l-2m}(\beta_R) 
\Gamma_l \Gamma_{l-2m} \EE
\BE {\cal Y}_{n',m'}(\beta_R) \equiv {2\over\pi^2}
\int_0^{\pi /2} d\phi~\int_0^\pi d\t~
\cos(2\phi n') \cos(\t m') ~\Theta(\t_c-|\phi|)Y(\phi,\t;\beta_R)~.\EE
\EML
On the plane $(\phi,\t)$, the function $Y(\phi,\t;\beta_R)$ has
a maximum on the line $\t=\pi/2$, which moves from the position 
$(\pi/2,\pi/2)$ at the beginning of renormalization procedure (when $\beta_R
\sim\beta_F$) towards the position $(0,\pi/2)$ when approaching the fixed
point $(\beta_R\to 0)$. Elsewhere, $Y(\phi,\t;\beta_R)$ is either quite flat,
or its contribution is eliminated by the cutoff factor $\Theta(\t_c-|\phi|)$
during the renormalization flow.
Therefore, we approximated the function $Y(\phi,\t;\beta_R)$ on the plane
$(\phi, \t )$ by its value on the line $(\phi,\pi/2)$. This
approximation, simplifying considerably our equations, allows an
analytical treatment and a qualitative insight
harder to find in purely numerical results. The approximate analytical
solution of the RG equations given below justifies that simplification {\it
a posteriori}, when compared with the direct numerical solution of
Eqs.~(\ref{RGangle}).

The approximate RG equations are:
\BML
\label{RGFNew}
\BE {\d\Gamma_n\over\d\beta_R} =~~
\sum_{m=-\infty}^\infty \bigg[ {1 \over \cosh^2\beta_R} \delta_{nm} 
-Y_{n-m}(\beta_R) \bigg] \Gamma_m^2 \EE
\BE {\d F_n\over\d\beta_R} = 
-\sum_{m=-\infty}^\infty Y_{n-m}(\beta_R) \Gamma_m^2~,\EE
\EML
wherein
\BE\label{FY}
Y_n(\beta_R) = {2 \over \pi} \int_0^{\arcsin(2 \beta_R/\beta_F)}
d\phi~ Y(\phi,{\pi\over 2}; \beta_R) \cos(2n \phi) ~.\EE
The key difference between Eqs.(\ref{RGRPA}) and (\ref{RGFNew}) is that
the former do not generate new harmonics since all harmonics are decoupled,
whereas the latter couple all harmonics (because of the $ZS'$ contribution)
in such a way that an infinite number of new harmonics are generated by the 
RG flow, even if only a finite number of harmonics are nonzero at the start.
For instance, the trial interaction (\ref{UbareF}) has only three nonzero
components, but according to Eqs.(\ref{RGFNew}) the fixed points $\Gamma^*$
and $F^*$ will possess an infinite number of them. The generation of new
harmonics is not an artefact of the approximation which was used to go from
Eqs.~(\ref{RGangle}) to Eqs.~(\ref{RGFNew}), but is a generic consequence of
the interference in the Landau channel (cf. Eqs.~(\ref{GFex})).

Let us start the analysis of Eqs.(\ref{RGFNew}) with a heuristic
observation. Whereas the component $Y_0(\beta_R)$ is a nonnegative function of
$\beta_R$, the others $(Y_n(\beta_R),~~n \geq 1)$ are increasingly
oscillating functions of $\beta_R$ when $n$ increases. These oscillations
along the whole RG trajectory $[0,\beta_0]$ will effectively decrease the
contributions from the harmonics $\Gamma_m$ $(m\neq n)$ to the flow of
$\Gamma_n$. Because of this, we expect the diagonal terms ($m=n$) of
Eqs.(\ref{RGFNew}) to be more important, and this justifies a perturbative
approach, in which the nondiagonal terms are ignored at zeroth order.
Let $\gamma_n(\beta_R)$ be the zeroth order solution:
\BE\label{QuasiRPA}
{\d \gamma_n \over \d \beta_R} 
= \bigg[ {1 \over \cosh^2\beta_R}- Y_0(\beta_R) \bigg] \gamma_n^2 ~.\EE
The solution is 
\BE\label{gamQRPA}
\gamma_n(\beta_R)= {U_n \over {1+ \bigg[\tanh\beta_0 -\tanh\beta_R 
-I_0(\beta_0)+ I_0(\beta_R) \bigg] U_n}}~,\EE
with 
\BEQ\label{In}
I_n(\beta_R) &\equiv& \int_0^{\beta_R} d\beta_R' ~Y_n(\beta_R') 
= {1\over\pi\beta_F}\int_0^{\arcsin(2 \beta_R/\beta_F)} d\phi\NN\\
&&\bigg[ \ln {\cosh(\beta_R + \beta_F\sin\phi) \over \cosh(\beta_R -
\beta_F\sin\phi) } - \ln {\cosh( \frac32 \beta_F \sin\phi) \over \cosh(
\frac12 \beta_F\sin\phi)}\bigg] {\cos(2n\phi) \over \sin\phi} ~.\EEQ
The fixed point $\gamma_n^*$ is
\BE\label{gamast}
\gamma_n^*= {U_n \over 1+ \bigg[1 -I_0(\beta_0) \bigg] U_n} ~.\EE
The integrals $I_n(\beta_0)$ can be evaluated analytically,
since $(\beta_F, \beta_0) \gg 1$ according to condition (\ref{TCond}).
In the following we shall need the first two components only:
\BML
\label{I0}
\BEQ I_0(\beta_0) &\approx& {\La_0 \over K_F}{1 \over \pi}
\left[\ln2 +\ln{1+\sqrt{1-\La_0^2/(2 K_F)^2}
\over 1+\sqrt{1-\La_0^2/K_F^2} } \right]\NN\\
&&+{1\over\pi}\bigg(2 \arcsin{\La_0 \over 2 K_F} 
-\arcsin{\La_0\over K_F} \bigg) + {T \over v_F K_F}
{(\ln2)(\ln3)\over\pi} \EEQ
\BEQ I_1(\beta_0) &\approx& {\La_0 \over K_F}{1 \over \pi}
\Big[\ln2 +\ln {1+\sqrt{1-\La_0^2/(2 K_F)^2}
\over 1+\sqrt{1-\La_0^2/K_F^2} }\NN\\
&&+\sqrt{1-\La_0^2/K_F^2} -
\sqrt{1-\La_0^2/(2 K_F)^2} \Big] 
+{T\over v_F K_F} {(\ln2)(\ln3)\over \pi}
~.\EEQ
\EML
(The next term in the temperature dependence, omitted in Eqs.~(\ref{I0}),
is of the order $(T/v_FK_F)^3$). Treating the off-diagonal terms
($n \neq m$) on the r.h.s. of (\ref{RGFNew}a) as
perturbations, we obtain the following approximate solution at first order:
\BML
\label{ApGam}
\BE
\Gamma_n(\beta_R) \approx \gamma_n(\beta_R) + \sum_{m \neq n}
\int_{\beta_R}^{\beta_0} d\beta_R' ~Y_{n-m}(\beta_R') 
\gamma_m^2 (\beta_R') \EE
\BE F_n(\beta_R) = \Gamma_n(\beta_R) + 
\int_{\beta_R}^{\beta_0} d\beta_R'\;
{\Gamma_n^2 (\beta_R')\over \cosh^2\beta_R'}~.\EE
\EML
It is straightforward to check that the solution
(\ref{ApGam}a) satisfies the sum rule (i.e., the Pauli principle
(\ref{Gamma0},\ref{GPauli})):
\BE\label{Gsumrule}
\sum_{n} \Gamma_n(\beta_R)=0~,~~\forall ~ \beta_R ~.\EE
The solution (\ref{ApGam}) can be converted back in terms of
the relative angle 
$\t\in [-\pi,\pi]$ with a little help from Eq.~(\ref{QuasiRPA}):
\BML
\label{Gamang}
\BEQ
\Gamma^*(\t) &=& U(\t) -
\int_0^{\beta_0} {d\beta_R \over \cosh^2\beta_R}~
\sum_{n=-\infty}^\infty \cos(n\t) \gamma_n^2 (\beta_R) \NN\\
&&+\Theta(\t_0-|\t|) 
\int_{\frac12 \beta_F |\sin(\t/2)|}^{\beta_0}d\beta_R~ 
Y({\t\over2},{\pi \over 2}; \beta_R)~
\sum_{n=-\infty}^\infty \cos(n\t) \gamma_n^2 (\beta_R) \EEQ
\BE F^*(\t) = \Gamma^*(\t) +
\int_0^{\beta_0} {d\beta_R \over \cosh^2\beta_R}~
\sum_{n=-\infty}^\infty \cos(n\t) \Gamma_n^2 (\beta_R) 
~,\EE
\EML
wherein $\t_0 \equiv 2 \arcsin(2 \beta_0/\beta_F)$. A comparison of
Eqs.~(\ref{Gamang}a) and (\ref{RGangle}a) shows that -- with the
aforementioned approximation of the angular dependence of the function $Y$ --
the approximate solution (\ref{ApGam}a) may be obtained by replacing the
vertex components $\Gamma_n$ on the r.h.s. of Eq.~(\ref{RGangle}a) by the
``renormalized'' RPA ansatz (\ref{gamQRPA}). It would be a mistake, however,
to conclude that the $ZS'$ diagram contributes only to the third term on the
r.h.s. of Eq.~(\ref{Gamang}a) since the $\gamma_n$-s partially include its
contribution. It is worth noting that Eqs.~(\ref{ApGam}b,\ref{Gamang}b) are
not approximations in the sense of Eqs.~(\ref{ApGam}a) or (\ref{Gamang}a),
but they are exact relations for $F$, derived from the basic RG
equations (\ref{RGFNew}). 

\subsection{Extension of the Effective Action}

In the numerical and analytical results presented in the following sections 
the initial cutoff $\La_0$ of the effective action is extended to $K_F$, i.e.,
$\beta_0 =\beta_F/2$. This point should be clarified. 
Notice first that the $ZS$ contribution is not sensitive to the bandwidth
cutoff $\La_0$ -- provided condition (\ref{TCond}) is satisfied -- since
$\tanh\beta_0$ is unity with exponential accuracy. On the other hand, the
angular cutoff of the $ZS'$ contribution (cf.
Eqs.~(\ref{RGangle},\ref{GFex},\ref{FY},\ref{In})) comes from a cutoff
imposed on the momentum transfer in this graph (cf. Eq.~(\ref{ZS'})). It is 
$\t_c=\arcsin(2\beta_R/\beta_F)$ (with $2\beta_R/\beta_F \equiv\La/K_F$) if
$\La_0 \leq K_F$, and $\t_c=\pi/2$ otherwise. The specific choice
$\beta_0 =\beta_F/2$ ($\La_0 = K_F$) means that at the initial point of the RG
flow the angle $\phi$ is allowed to take all values (i.e., the momentum
transfer ${\bf Q}'$ is not cut off), while the bandwidth is extended to the
full depth of the Fermi sea. It can be checked that the results are not
sensitive to the choice of a bigger cutoff $\La_0\gtrsim K_F$, since
then not only is the $ZS$ contribution to the flow is exponentially small, but
that of $ZS'$ as well, until the cutoff decreases to $\La \sim K_F$ (this was
also confirmed by direct numerical tests). The formulas for the approximate
analytic solution are derived for $\Lambda_0 \leq K_F$.

Such an extension of the low-energy cutoff to large values is analogous to
what is routinely done in $1D$ models (e.g., the Tomonaga-Luttinger
model\cite{Voit}). In that context, deviations of the real excitation spectrum
from linearity and the approximated integration measure are expected to affect
only the numerical values of the renormalized physical parameters. 

Choosing $\Lambda_0 \sim K_F$ renders the RG fixed points
(observables) sensitive only to the two independent physical
scales present in the model: $T$ and $v_FK_F=2E_F$, and not to the 
arbitrary scale $\La_0$, which divides fast and slow modes.
Lowering the running cutoff until it reaches some intermediate scale
$\Lambda_X$ (such that $\Lambda_X\ll K_F$ and $v_F\Lambda_X\gg T$) provides us
with $\Lambda_X$-dependent parameters for the action. We regard $\La_X$ as the
scale of the low-energy effective action. However, the observable quantities
(the fixed points) do not depend on a particular choice of $\La_X$.

\section{Analysis and Discussion of the RG Results}
\label{discussionS}

We will now discuss the main novelties brought by quantum interference in the
Landau channel and compare with the results of decoupled approximations.
The solutions $\Gamma^*(\t)$ and $F^*(\t)$ at different temperatures and for
the interaction (\ref{UbareF}) are shown on Fig.3 (A: direct numerical
solution of Eqs.~(\ref{RGangle}); B: solution (\ref{Gamang})). 
For this interaction the sum in the
second and third terms on the r.h.s. of Eq.~(\ref{Gamang}a) is
$\gamma_0^2 (\beta_R)+2 \gamma_1^2 (\beta_R)\cos\t$.
The curves
were calculated for ${\cal U}=1$ (cf. Eq.~(\ref{Ubare})), which is four times
smaller than the critical value ${\cal U}_{cr}^{\rm RPA}=4$ at which the
instability appears in the RPA solution (\ref{RGRPA}a) for $\Gamma_1^*$.
Comparison of the approximate solutions (\ref{ApGam},\ref{Gamang}) with the
direct numerical solution shows good agreement.

In Fig.~3 the differences between the RG solution and the RPA solution
(\ref{RGRPA}a) are minor at large angles, but they become especially striking
at small angles $\t$, where the interference between the $ZS$ and 
$ZS'$ contributions is very strong. The RG solution gives 
$\Gamma^*(\t=0)=0$ (the Pauli principle), while $\Gamma^{\rm RPA}(\t=0)=-1/3$
for this interaction strength. The Landau interaction function 
$F^*(\t)$ differs from the bare interaction $U(\t)$, and
$F^*(\t=0) \neq 0$. If the $ZS'$ contribution is neglected (the RPA solution
(\ref{RGRPA}b)), these two quantities coincide.

An interesting feature of the RG result is the temperature dependence of the
vertices $\Gamma^*(\t)$ and $F^*(\t)$. As $T$ decreases, the
``beak'' of $\Gamma^*(\t)$ in the region of strong
interference becomes narrower. The characteristic angular width of this
``beak'' is $|\t| \sim T/v_FK_F$. A similar narrowing is noticeable in the
temperature dependence of $F^*(\t)$. One can also see from the figures a
weakening of the interference effect at lower temperatures, for then the RG
solutions lie closer to the RPA curves, but the distinctions between them do
not disappear as $T \to 0$, and the RG never reproduces the RPA
result.\cite{noteT0} 

In terms of Fourier components this behavior manifests itself in a linear
temperature dependence of $\Gamma^*_n$ and $F^*_n$. This linearity is found
both in the direct numerical solution of Eqs.~(\ref{RGangle}), and from the 
solution of Eqs.(\ref{FY},\ref{gamQRPA},\ref{In},\ref{ApGam}). This
temperature dependence can be revealed analytically. Integrating by parts and
using Eq.~(\ref{QuasiRPA}), we can rewrite Eq.~(\ref{ApGam}a) at the fixed
point as
\BE\label{GBP}
\Gamma_n^* =\gamma_n^* + \sum_{m \neq n}I_{n-m}(\beta_0)U_m^2 -2
\int_0^{\beta_0} d\beta_R \sum_{m \neq n} I_{n-m}(\beta_R)
 \bigg[ {1 \over \cosh^2\beta_R} - Y_0(\beta_R) \bigg] \gamma_m^3(\beta_R)
~.\EE
The leading term on the r.h.s. of Eq.~(\ref{GBP}) is
$\gamma_n^*$. Using then Eq.~(\ref{gamast},\ref{I0}a), we obtain for
$n=0,1$:
\BE
\label{Gamas}
\Gamma_n^*(T) \approx \gamma_n^*(T)
\approx {\gamma}_n^*(0) + {T \over v_F K_F} {(\ln2)(\ln3)\over \pi}
[{\gamma}_n^*(0)]^2~,~(n=0,1)~,\EE
wherein 
\BE\label{gamT0} {\gamma}_n^*(0) = {U_n \over {1+ \bigg[1 - I_0(\beta_0)
\bigg|_{T=0} \bigg] U_n }} ~.\EE
For the interaction (\ref{UbareF}) $U_n=0$ and so
$\gamma_n^*=0$ for $n>1$. Thus, the higher harmonics $\Gamma_{n>1}^*$,
are entirely generated by the RG flow. To leading order, we obtain from
Eq.~(\ref{GBP}):
\BE\label{Gam2}
\Gamma_2^* \approx I_1(\beta_0)U_1^2 ~.\EE
This component also has a linear temperature dependence,
according to Eqs.~(\ref{I0}). To estimate the components of the Landau
function, we first rewrite Eq.~(\ref{ApGam}b) in another, equivalent form (cf.
Eqs.~(\ref{RGFNew})):
\BE\label{Feq} F_n(\beta_R) = U_n + \sum_{m=-\infty}^\infty 
\int_{\beta_R}^{\beta_0} d\beta_R' ~Y_{n-m}(\beta_R') 
\Gamma_m^2 (\beta_R') ~.\EE
Proceeding in the same fashion as above, we obtain the
linear temperature-dependent components $F_n^*$:
\BML\label{Fnas}
\BE
F_n^* \approx U_n +I_0(\beta_0)U_n^2+
(|n-1|+1)I_1(\beta_0)U_{|n-1|}^2 , ~~(n=0,1) \EE
\BE F_2^* \approx I_1(\beta_0)U_1^2 ~.\EE
\EML
We should emphasize that simple formulas like
(\ref{Gamas},\ref{Gam2},\ref{Fnas}) serve only to illustrate how the
temperature dependence comes about, and give only the order of magnitude of
the higher harmonics ($n > 2$). The latter should rather be calculated
numerically. The temperature dependence of the lowest harmonics (e.g.,
$F^*_0$ and $F^*_1$) does not seem to be a relevant issue in the calculation
of quantities such as the compressibility, effective mass and heat capacity, 
since, in the total $ZS'$ contribution, the temperature corrections, of the
order of $T/v_FK_F$, are very small in comparison with the main corrections of
order $\Lambda_0 /K_F$. As a consequence, the actual values
of the lowest harmonics vary within a few percent at most, even in the entire
temperature interval $0 \leq T/v_FK_F \leq 0.1$ (the maximum temperature
studied is really high: $T=0.2 E_F$).

The temperature dependence is more pertinent as a ``collective'' effect of
the higher harmonics generated by the RG flow. Let us explain
this point with the example of the interaction (\ref{UbareF}). The ``improved''
RPA ansatz (\ref{gamQRPA}) renormalizes the bare components $U_n$ into
$\gamma_n$ ($n=0, \pm 1$). The latter form almost perfectly the function
$\Gamma^*(\t)$, except at small angles. For those three components
$\gamma_n$ the sum rule (\ref{Gsumrule}) is less violated than for the
``pure'' RPA components (\ref{RPAFP}a). The generation of the new harmonics
by the second term on the r.h.s. of Eq.~(\ref{ApGam}a) gives ``a final
touch'' to the curve $\Gamma^*(\t)$, resulting mostly in the formation of a
temperature-dependent feature near $\t=0$. The actual calculation of the
components $\Gamma^*_n$ showed that, in order to obtain with acceptable
accuracy the right form of $\Gamma^*(\t)$ provided by Eq.~(\ref{Gamang}a) via
the Fourier transformation of Eq.~(\ref{ApGam}a), at least
$N_{\rm max} \sim v_FK_F/T$ components are necessary. So, the lower the
temperature is, the more harmonics are needed for the formation of the vertex
$\Gamma^*(\t)$. The same conclusion can be drawn from a numerical solution of
the equations, but since it is carried out in terms of angles on a discrete
grid, a reliable calculation of higher harmonics is difficult.

Another physical consequence of the quantum interference in the Landau
channel is the increased robustness of the system against instabilities
induced by strong interactions. 
Even from the approximate solution~(\ref{ApGam}), we see that the maximum
interaction strength allowed is now larger than the one provided
by the RPA solution (cf. (\ref{RPAFP},\ref{Stab1})).
From Eq.~(\ref{gamast}) we obtain the stability conditions
for the approximate solution (\ref{ApGam}): $U_l >
-[1-I_0(\beta_0)]^{-1}, ~\forall~ l$ with $0<I_0(\beta_0)<1$ according to
(\ref{I0}a). Since $I_0(\beta_0)$ grows with temperature, larger values of
$|U_l|$ are allowed as $T$ increases: the higher the temperature, the more
stable the system is, as it should be from physical grounds.
At the optimal choice of the initial cutoff ($\La_0 = K_F$),
$I_0(\beta_0)$ grows from 0.255 at
$T=0$ to 0.27 at $T/v_FK_F=0.1$. This value of temperature is the largest we
can try without violating the condition of applicability of our model
(\ref{TCond}). Thus, within this approximate solution, the effect of
interference increases the critical coupling by 40\% compared the the RPA
critical value (\ref{Stab1}). Since we are retaining only two one-loop
diagrams, linearized excitation spectrum and integration measure, we cannot
be more conclusive on the role of the modes deep into the Fermi sea in
screening a microscopic interaction of arbitrary strength, and in stabilizing
the Fermi liquid phase.

\section{Contact with the Landau FLT and Discussion}
\label{contactS}

In this section we explain how the present RG theory is related to the
standard results of the Landau FLT.\cite{Landau57,Landau59} This will also
allow us to relate this study to previous work on this RG approach to
the Fermi liquid.\cite{GD95,NG96}

It is important to notice that the two contributions to the RG flow, coming
from the $ZS$ and $ZS'$ graphs, behave quite differently as the flow
parameter $\beta_R$ runs from $\beta_0 \gg 1$ towards $\beta_R=0$. At large 
$\beta_R$ the $ZS$ contribution to the flow, which gives the term
proportional to $\cosh^{-2} \beta_R$ on the r.h.s. of Eq.~(\ref{GFex}), is
virtually negligible, up to $\beta_R \sim 1$. On this part of the RG
trajectory, the main contribution to the renormalization of
$\Gamma$ and $F$ comes from the $ZS'$ graph. On the other hand, closer to the
fixed point ($\beta_R \lesssim 1$), the $ZS$ contribution grows since
$\cosh^{-2}\beta_R \sim 1$ for all harmonics, while $Y_n(\beta_R)$ decreases
for the lower-order harmonics. At $\beta_R \ll 1$:
\BE\label{Yas2} Y_n(\beta_R) \approx {1 \over \pi n} \sin {4n \beta_R \over
\beta_F} ~.\EE 
Using the approximated form (\ref{FY}) is justified here, since
at $\beta_R\ll 1$ there is no difference between the exact form of the RG
equations (\ref{GFex}) and Eqs.~(\ref{RGFNew}). Indeed, when $\beta_R \ll 1$, 
the largest allowed $\phi$ is roughly $ 2 \beta_R / \beta_F$, so in
Eq.(\ref{Ybetax}) 
${\rm max} | \beta_{Q^\prime} | \approx 2 \beta_R \ll 1$ and the limit 
(\ref{limY}) of the function $Y$ can be taken. The Kronecker delta appearing 
after the integration over 
$\t$ removes one summation, and we recover exactly Eqs.~(\ref{RGFNew})
with $Y_n(\beta_R)$ given by (\ref{Yas2}). It should be also kept in mind
that the $ZS'$ flow is localized within the angle $|\phi| \sim 2 \beta_R /
\beta_F$.

Such different behavior of the two contributions ($ZS$ and $ZS'$) to the total
RG flow explains why approximations based on the decoupling of these two 
contributions (RPA, $ZS$-ladder\cite{AGD,Lifshitz80,GD95,NG96}) are
reasonable. To clarify to what extent the standard results of
FLT (\ref{RPALandau},\ref{SRLandau}) can be corroborated by RG, we will
make a two-step approximation of our RG equations. In doing so we will follow
exactly the ``recipe'' of the $ZS$-ladder approximation discussed in Sec.V,
but now we can check each step by direct comparison with the RG solution of
Eqs.~(\ref{RGangle}).

In the first step we neglect the contribution of the $ZS$ graph above an
intermediate flow parameter $\beta_X$. As one can see from the RG equations
(\ref{GFex}), this removes the exponentially small difference between 
$\Gamma_n( \beta_R)$ and
$F_n( \beta_R)$ at $\beta_R > \beta_X$. This approximation is asymptotically
exact as $T \to 0$.\cite{note0} Neglecting, in the second stage of this
approximation, the $ZS'$ flow for $\beta_R < \beta_X$, localized by that time 
within the angle $\t_X = 2 \arcsin(2 \beta_X / \beta_F)$, we recover the
exactly solvable equations (\ref{RGRPA}) with the new initial point $\beta_R =
\beta_X$, instead of $\beta_R = \beta_0$. Then according to
Eqs.~(\ref{RGRPA}), $F_n^{X} \equiv F_n(\beta_X)$ is the (approximate) fixed
point value of the Landau function, while $\Gamma_n(\beta_R)$ flows towards
the (approximate) fixed point $\Gamma_n^{\rm ph}$ from the new bare value
$\Gamma_n^X \equiv \Gamma_n(\beta_X) = F_n^{X}$. This second step of
approximation violates the Pauli principle, no matter how close we are
to the Fermi surface (cf. Eq.(\ref{Gamma0}) and Ref.\onlinecite{note0}).
Afterwards the theory says nothing about the values of the functions
$\Gamma(\t)$ and $F(\t)$ inside the interval $2 \t_X$ and, of course, there
are no more correlations between these functions. 
 
To preserve the correct zero-temperature limit and to minimize the angle 
within which the approximation gives completely wrong results for $\Gamma^*$ 
and $F^*$, the intermediate cutoff $\La_X$ corresponding to
$\beta_X=v_F \La_X /2T$ should be chosen such that $\tanh \beta_X \approx
1$ (cf. Eqs.~(\ref{RGRPA},\ref{RPAFP}) and Ref.\onlinecite{GD95})
and $2\beta_X / \beta_F =\La_X / K_F \ll 1$. Summing up what is said
above, we obtain:
\BML
\label{Lappr}
\BE\Gamma_n^{\rm ph} = {\Gamma_n^X \over {1+ \tanh(\beta_X) \Gamma_n^X } }
= {F_n^{X} \over {1+F_n^{X}} } \EE
\BE \Gamma_n^X=F_n^{X} = U_n +
\sum_{l, m=-\infty}^\infty \int_{\beta_X}^{\beta_0} d\beta_R' ~
{\cal Y}_{n-m, 2l-2m}( \beta_R') \Gamma_l (\beta_R')
\Gamma_{l-2m}(\beta_R')~.\EE
\EML

In Fig.~4 we illustrated all this by the direct numerical calculation of
$F^{X}$, $\Gamma_n^X$ from Eqs.~(\ref{RGangle}) for the interaction
(\ref{Ubare}), followed by
a calculation of $\Gamma^{\rm ph}$ from Eqs.~(\ref{Lappr}). The RG solutions
for $\Gamma^*$ and $F^*$ are also presented. The function
$F^{X}(\t)$ follows almost perfectly the Landau function (the real fixed point
$F^*(\t)$), except within $2 \t_X$ of $\t=0$.
In the part of the RG
trajectory $\beta_X \leq \beta_R \leq \beta_0$ ($\beta_0=100$, $\beta_X=5$,
$T/v_FK_F =0.005$), not only is the $ZS$ flow exponentially weak, but the
central part of the $ZS'$ flow as well (cf. Eq.~(\ref{limY})). So, the
evolution of both vertices is due mostly to the ``tail'' $\t > \t_X$ of the
function $Y$ at $\beta_R \gtrsim 1$. 
That is why $\Gamma^X(\t)$ and $F^{X}(\t)$ are virtually identical. 
Only the slowing down 
of the $ZS'$ flow almost everywhere at $\beta_R \lesssim 1$ -- except on the
central part (cf. Eq.~(\ref{Yas2})) wherein it is always as strong as the
other one ($ZS$) -- results in the drastic differences between the two limits
of the four-point vertex at the fixed point. The function $\Gamma^{\rm
ph}(\t)$ is featureless and looks like a corrected RPA solution. The
differences between
$\Gamma_n^*$ and $\Gamma_n^{X}$ ($F_n^*$ and $F_n^{X}$) are negligible,
i.e. less than 1\%, only for the components $n=0,1$.

As it should be clear by now, there is no real incompatibility
of the stability conditions with the Pauli principle, since this is a mere
artefact of the ZS-ladder approximation. It is pointless to impose the sum
rule either to
$\Gamma_n^{\rm ph}$ in the form (\ref{Gsumrule}), or to $F_n^{X}$ in the form
(\ref{SRLandau}). Both sums would give the value of the ``uncorrelated''
function $\Gamma^{\rm ph}(\t)$ at $\t =0$. This function goes smoothly
from the right patch $[\t_X, \pi]$ towards $\t=0$ (cf. Fig.~4) -- or,
equivalently, from the left, because of parity . Actually, it can be proved
exactly, turning the arguments of Sec.V around, that in a stable Fermi
liquid, it is impossible to obtain $\Gamma^{\rm ph}(\t=0)=0$, even by
chance. Thus, there is no need for the Landau function $F^*$ to be ``fine
tuned'' in the sense of the sum rule (\ref{SRLandau}), since only 
the relation (\ref{Lappr}) -- between the approximate vertex 
$\Gamma^{\rm ph}$ and $F^X$ --
is an exact relationship (more precisely,
asymptotically exact when $T\to 0$), not (\ref{RPALandau}), which
relates the physical quantities $F^*$ and $\Gamma^*$.

In the context of our discussion at the end of Sect.~\ref{solutionS}, notice
that the cutoff $\La_X$ ($v_F \La_X/T \gg 1$, 
$ \La_X/K_F \ll 1$) corresponds to the initial cutoff of the 
{\it low-energy} effective action wherein $\Gamma^X$ is the bare interaction 
function (coupling) of that action.
The equality of the functions $\Gamma^{X}$ and $F^{X}$ illustrates the 
point of Sect.~\ref{couplingS} that, at the beginning, the action's coupling 
function can be defined independently of the order in which 
the zero-transfer limit is taken.

When the RG flow reaches the scale $\La_X$, the contribution of the
$ZS'$ graph to the flow of $\Gamma_n$ and $F_n$ is strictly irrelevant in
the RG sense, and could have been neglected in a model with a finite number
of couplings (e.g., the $\varphi ^4$ theory, 1D g-ology models, and so on),
keeping only marginal terms (cf. Eqs.~(\ref{RGRPA})). But, as
pointed out by Shankar,\cite{Shankar94} in the vicinity of the Fermi surface we
are dealing with coupling {\it functions}, i.e., with an {\it infinite} set of
couplings. Our RG solution provides a curious example of a finite deviation
of the RG trajectory at the fixed point due to an infinite number of
irrelevant terms. The {\it right} fixed point ($\Gamma^{*}(\theta =0)=0$)
cannot be reached if those terms are neglected, since 
$\Gamma^{\rm ph}(\t_X\to0) \ne \Gamma^*(\t=0)$ (even at $T=0$\cite{noteT0})
and we would return to the
problems caused by the solution $\Gamma^{\rm ph}$ (the $ZS$-ladder
approximation) discussed in Sec.V. To put it differently, neglecting those
irrelevant terms at some part of the flow (solution (\ref{Lappr})) violates
the invariance of the RG trajectory at the point $\theta=0$, expressed
by Eqs.~(\ref{Gamma0},\ref{Init},\ref{GPauli}).

The $ZS$-ladder approximation seems acceptable in the 
normal Fermi liquid regime with moderate interaction ($F_n \lesssim 
10$), when the narrow-angle features of vertices revealed by the RG theory are
not too large,\cite{noteT0} because the forward ($\t=0$) singularity 
has little effect on the first components ($\Gamma_n^* \approx \Gamma_n^{\rm
ph}$, $F_n^* \approx F_n^{X}$ for $n=0,1$ and, in the case of a weak
interaction, for $n=2$). This singularity affects mostly the higher Fourier
components. So, the relationship (\ref{RPALandau}) is valid only for small
$n$. It should not be used for $F_n^*$
($n \geq 2$) neither directly, nor via the sum rule from the scattering
vertex provided experimentally. For the physical vertex $\Gamma_n^*$ the sum
rule (\ref{Gsumrule}) is always valid, but this study indicates that its
angular shape may require a large number of harmonics to adequately represent
it. The existence of a finite solution for $\Gamma^{\rm ph}(\t)$ under
conditions
\BE\label{stabNum}\Gamma_n^X > -1,~ \forall n \EE
guarantees not only finite RG solutions for $\Gamma^*$ and $F^*$, but
also the fulfillment of the thermodynamic Pomeranchuk conditions
(\ref{StabPom}) by $F^*$. 

The major consequence of this study on the standard results of the Landau FLT
is reducing the relationship (\ref{RPALandau}) between the components of the 
scattering vertex and the Landau function to the rank of approximation and
invalidating of the sum rule (\ref{SRLandau}). The rest of results for normal
Fermi liquids would not be affected seriously by the RG corrections. For
example, the temperature dependence of the vertices would give a weak
correction to the leading terms. These conclusions are neither related to the 
specific choice model considered, nor to the spatial dimension. Including
spin doubles the number of vertices involved, changing nothing essentially.
(The derivation of the RG equations with spin is straightforward using the
$N$-flavor formalism of Ref \onlinecite{GD95}.) The differences for the
case $d=3$ are only quantitative (e.g., the type of the temperature
dependence) because of different angular functions and solid angle
integrations.

\section{Summary}

In studying the Fermi-liquid regime of interacting fermions in $d>1$ with the
model of the $\psi^4$-Grassmann effective action as starting point of the
analysis, one must distinguish between three quantities: ({\it i}) the bare
interaction function of the effective action; ({\it ii}) the Landau
interaction function; ({\it iii}) the forward scattering vertex. We have
derived the RG equations for the Landau channel which take into account both
contributions of the $ZS$ and $ZS'$ graphs at one-loop level. The basic
quantities of the Fermi liquid theory, the Landau function and the scattering
vertex, are calculated as fixed points of the RG flow in terms of effective
action's interaction function. 

The classic derivation of Fermi liquid theory using the Bethe-Salpeter
equation for the four-point vertex at $T=0$ is based on the approximation that 
the vertex irreducible in the direct particle-hole loop ($ZS$) is a regular
function of its variables, neglecting the zero-angle singularity in the
exchange loop ($ZS'$). This approach is equivalent to our earlier decoupled RG 
approximation\cite{GD95,NG96}, and they are both tantamount to summation of
the direct particle-hole ladder diagrams, wherein the Landau function stands
as the bare interaction (the $ZS$-ladder approximation).

One of the major deficiencies of the $ZS$-ladder approximation is that the
antisymmetry of the forward scattering vertex related by the RPA-type formula
to the Landau interaction function, is not guaranteed in the final result,
and the amplitude sum rule must be imposed by hand on the components of the
Landau function. This sum rule, not indispensable in the original
phenomenological formulation of the Landau FLT\cite{Landau57}, from the RG
point of view is equivalent to fine tuning of the effective interaction. 
 
The strong interference of the direct and exchange processes of the
particle-hole scattering near zero angle invalidates the $ZS$-ladder
approximation in this region, resulting in temperature-dependent narrow-angle
anomalies in the Landau function and scattering vertex, revealed by the RG
analysis. In the present RG approach the Pauli principle is automatically
satisfied. As follows from the RG solution, the amplitude sum rule being an
artefact of the $ZS$-ladder approximation, is not needed to respect 
statistics and, moreover, is not valid.

\acknowledgements

Stimulating conversations with C.~Bourbonnais, N.~Dupuis and
A.-M.~Tremblay are gratefully acknowledged. In particular we thank
A.-M.~Tremblay for careful reading of the manuscript. This work is supported
by NSERC and by F.C.A.R. (le Fonds pour la Formation de Chercheurs et
l'Aide \`a la Recherche du Gouvernement du Qu\'ebec).

%

\begin{figure}
\vglue 0.4cm\epsfxsize 6cm\centerline{\epsfbox{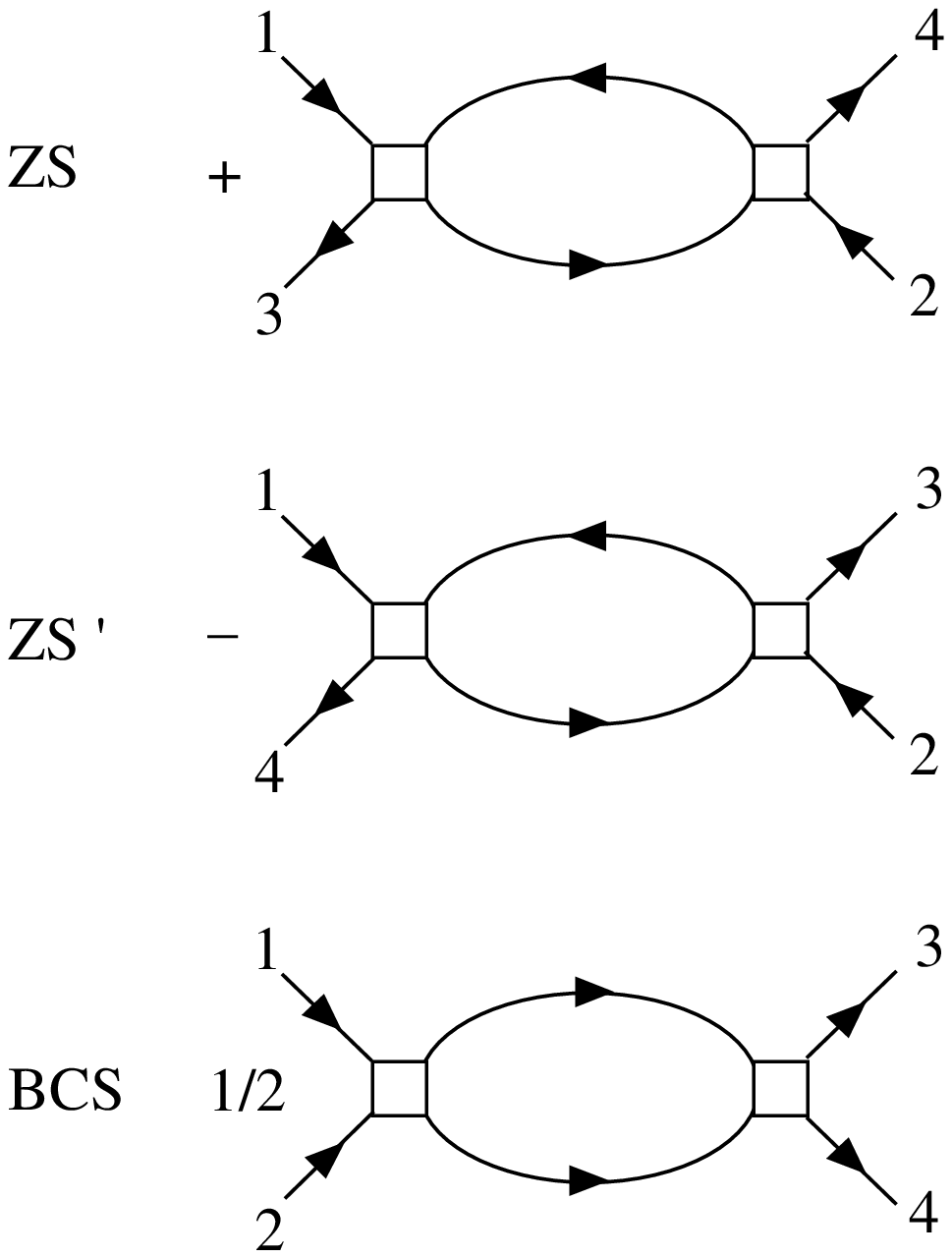}}\vglue 0.4cm
\caption{\baselineskip=12pt
The three diagrams contributing to the RG flow at one-loop.
}
\end{figure}
\begin{figure}
\vglue 0.4cm\epsfxsize 12cm\centerline{\epsfbox{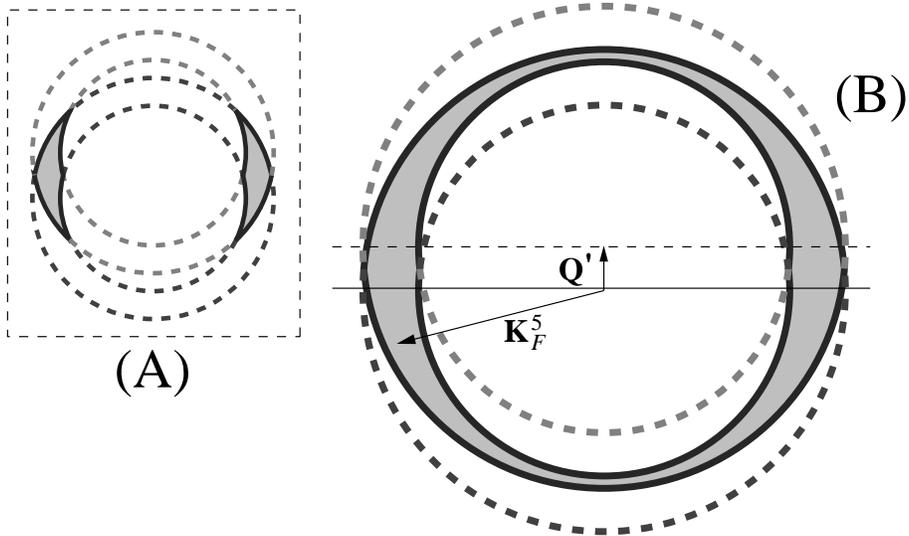}}\vglue 0.4cm
\caption{\baselineskip=12pt
If $|{\bf Q}'|>2\La$, the intersection (shaded) of the supports of ${\bf
K}_5$ and ${\bf K}_5-{\bf Q}'$ are disconnected (A). If
$|{\bf Q}'|<2\La$, this intersection forms a connected area (B). Note
that the RG flow is governed by the boundaries of this intersection, not by
their interior directly.
}
\end{figure}
\begin{figure}
\vglue 0.4cm\epsfxsize 12cm\centerline{\epsfbox{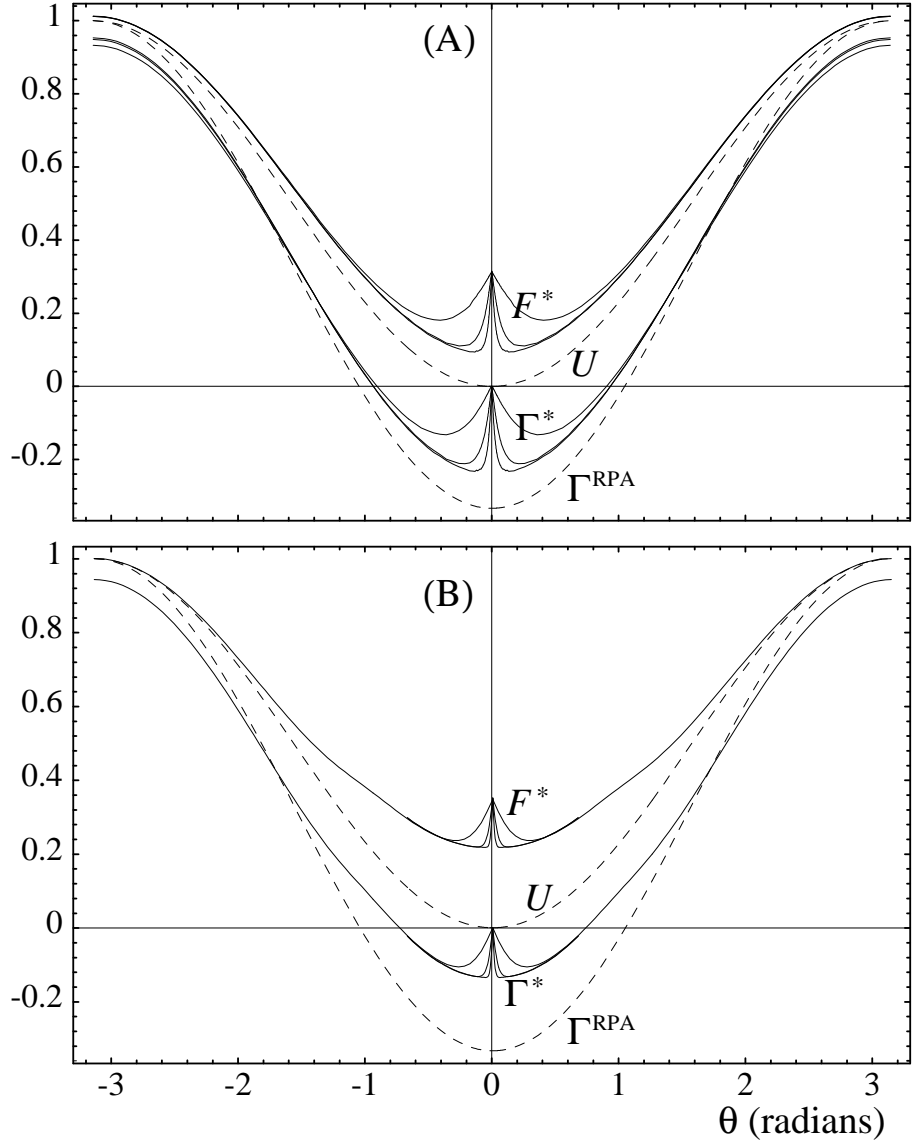}}\vglue 0.4cm
\caption{\baselineskip=12pt
({\bf A}) Results of the numerical solution of the coupled RG equations. The
curves labeled $\Gamma^*$ and $F^*$ are the forward scattering vertex and
the Landau function, respectively, at temperatures $T/v_F K_F =$ 0.1, 0.025, 
and 0.01. The narrowest central peak corresponds to the smallest
temperature, and vice versa. ({\bf B}) Approximate analytical solution of the
coupled RG equations, for the same parameters as in (A), calculated
numerically from Eq.~(\ref{Gamang}). In both cases the initial cutoff was
$\La_0=K_F$. }
\end{figure}
\begin{figure}
\vglue 0.4cm\epsfxsize 12cm\centerline{\epsfbox{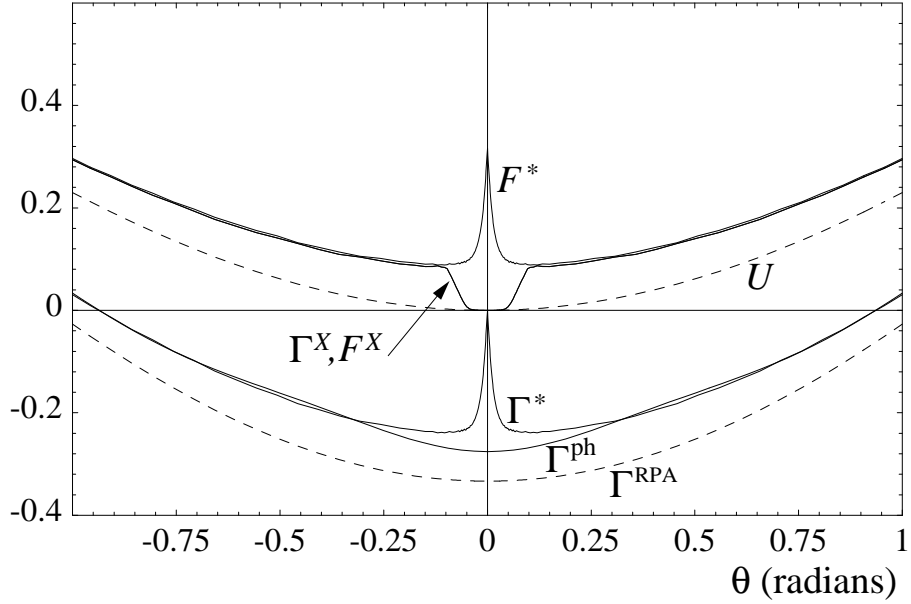}}\vglue 0.4cm
\caption{\baselineskip=12pt
Comparison between the exact numerical solution of the coupled RG equations
for $T/v_F K_F=0.005$ ($\Gamma^*$ and $F^*$), the intermediate values of
$\Gamma^X, F^X$ obtained from the initial value $U$ by stopping the
flow at $\beta_R=5$, and the phenomenological vertex $\Gamma^{\rm ph}$ 
(the result of the standard FLT derivations) obtained
by applying the RPA solution to $\Gamma^X$ ($F^X$) considered as a new initial 
point of the flow. $\Gamma^{\rm ph}$ practically coincides with $\Gamma^*$, 
except in the central region.}
\end{figure}


\begin{references}
\bibitem{note0} Both the $ZS$ contribution and the zero-angular part of the
$ZS'$ contribution become singular in the limit $T \to 0$ since 
$\lim_{\beta\to \infty } 
(\beta /4)\cosh^{-2}(\beta x/2)=\delta (x)$.
\bibitem{noteT0}
At exactly zero temperature the angular features of the vertices reduce 
to finite discontinuities at 
zero angle, accompanied by finite-angle deviations from the RPA curve
depending on the parameters of the effective action, e.g., the radius of Fermi 
surface, the strength of the interaction, etc. The zero-temperature limit is, 
however, mostly an academic question since the effect of interference 
between the Landau and the BCS channels, neglected in this study, would result 
in a Kohn-Luttinger instability in the $V$-interaction function\cite{Shankar94} 
and destroy the regime of Fermi liquid before the system attains $T=0$. Due 
to this second interference our results are not reliable near $\theta =\pi$, 
since $\Gamma( \pi)=V(0)$.\cite{Shankar94,GD95}
\end{references}
\end{document}